**Data-driven modeling in the introductory physics laboratory: Scaling analysis and data collapse in the specific heat of water experiment**

Kazumasa Kushida[1]*, Tomohiro Oda[2], Koji Yamaguchi[3]

[1] Department of Educational Collaboration, Osaka Kyoiku University, Kashiwara, Osaka 582-8582, Japan

[2] Ikeda Senior High School Attached to Osaka Kyoiku University, Ikeda, Osaka 563-0026, Japan

[3] Tennoji Senior High School Attached to Osaka Kyoiku University, Tennoji, Osaka 543-0054, Japan

## Abstract

We present a data-driven redesign of the classic specific-heat-of-water experiment that uses scaling analysis and data collapse for model construction rather than beginning with a previously established thermodynamic formula. The activity combines structured experimental and analytical guidance with instructor-mediated questioning, while the thermodynamic description is deliberately postponed. Students collect temperature–time data under different conditions and obtain a collapsed relation between the mass-normalized energy input and temperature rise through successive rescaling. They formulate an empirical model from this relation and test it using additional data not included in its construction. In the post-experimental discussion, the thermodynamic description shows that the fitted empirical constant characterizes the combined response of the water and calorimeter and approximates a quantity that depends weakly on water mass. Reanalysis of the mass-dependent data separates the two contributions, with limited precision over the available water-mass range. The copper cup is an intentional feature of the instructional design: its contribution is large enough to require physical interpretation but varies little enough over the explored mass range for the data to collapse approximately onto a single relation. The same apparatus thereby supports empirical model construction, thermodynamic interpretation, and separation of the component contributions.

* *Corresponding Author*: Kazumasa Kushida
* *E-mail*: kkushi@cc.osaka-kyoiku.ac.jp

## I. INTRODUCTION

Constructing and evaluating models is a central practice of physics.[1] Modeling practices include identifying relevant variables, formulating relationships among them, and judging the domain over which the relationships remain valid. Graphical analysis is widely used in introductory laboratories to extract physical quantities from linearized forms of established relations.[2] Other laboratory curricula have students design experiments and construct relationships from experimental data before the corresponding theory is introduced.[3] At the upper-division level, model construction, prediction, data interpretation, and model refinement have been made explicit learning goals of a laboratory activity.[4]

In the present activity, an introductory specific-heat laboratory, students likewise construct an empirical model from experimental data before the corresponding theory is introduced. With structured experimental and analytical guidance, they obtain the model through successive scaling and data collapse without invoking thermodynamic formulas. The copper cup is retained as an intentional feature of the instructional design: its thermal contribution is not revealed to students in advance but becomes a subject of analysis after the empirical model has been constructed and tested. In the post-experimental discussion, students examine what physical system the model describes, and the water and calorimeter contributions are separated.

The outline of this paper is as follows. Section II describes the apparatus and instructional design. Section III presents the data-driven scaling analysis, empirical model, and predictive test. Section IV introduces the thermodynamic description, clarifies the physical meaning of the parameter in the empirical model, and uses the mass-dependent data to separate the water and calorimeter contributions. Section V presents the pedagogical implications, discussion, and conclusions.

## II. EXPERIMENTAL CONTEXT AND DESIGN PHILOSOPHY

### A. Experimental system and measurement conditions

The activity has been implemented in an undergraduate introductory physics laboratory course, with students working in groups of two or three over two 90-minute sessions. The experimental system consists of a simple calorimeter and a digital thermometer. The calorimeter (WKA, KENIS LIMITED, Osaka, Japan) contains a 110 g copper cup surrounded by 1.5 cm of expanded polystyrene within a plastic outer vessel (Fig. S1 of the Supplementary Material). A coiled nichrome-wire heater supplied with the calorimeter is operated using a regulated dc power supply (KP-5A, KENIS) at currents of 2.4–4.0 A and voltages of 3.7–6.2 V. Temperature is measured using a digital thermometer (CT-450WR, CUSTOM Corporation, Tokyo, Japan) with a display resolution of 0.1 °C. The use of the copper cup is an intentional feature of the instructional design, as discussed in Sec. V. Further apparatus and procedural details, including the purchase price, are provided in Sec. S2.1 of the Supplementary Material.

The power supplied to the heater is kept constant during each run and is determined from the measured current and voltage. For readability, heater powers are shown as nominal values rounded to the nearest watt in the text and figures, whereas all calculations use the recorded products of current and voltage. The temperature increase $\Delta T$ is recorded as a function of heating time. Measurements are limited to a temperature increase of approximately 5 K to reduce the effects of temperature-dependent heat capacities, convection, and radiation; no corrections for these effects are applied. Our aim is not high-precision calorimetry, but rather to produce trends that can be compared and rescaled across different experimental conditions. To ensure uniform heating and reliable temperature measurements, the heater and thermometer must remain fully immersed, limiting the practical range of water mass. In the implementation described here, water masses range from 170 to 240 g and heater powers from 9 to 25 W. Excessive nonlinearity in students' plots usually indicates insufficient mixing, unintended contact among the heater, thermometer, and stirrer, or poor placement of the thermometer.

**B. Black-box framing and instructional guidance**

In a study of two introductory physics courses, students who did and did not take an associated laboratory course designed to support the lecture content showed no measurable difference in performance on related final-examination questions.[5] The present activity instead assigns the laboratory a different role: students first establish an empirical relationship from the experiment itself, and theory is introduced only after that relationship has been constructed and tested. To implement this role, students are instructed during the initial stages of data analysis to treat the system as a black box whose internal thermal structure is not specified in advance. Within this framing, the modeling is data-driven: until the post-experimental discussion, students propose and evaluate an empirical model based on relationships observed in the measured data without using thermodynamic equations to determine the form of the model. This black-box framing should not be interpreted as unguided discovery. Students are provided with instructions for operating the apparatus, recording data, selecting experimental conditions within specified ranges, and organizing their results for analysis.

Students use their data to infer relationships among heating time, electrical power, water mass, and temperature increase and then explore these relationships through graphical analysis and rescaling. The analytical process is supported by instructional materials that outline the main sequence of measurements, graphing, and rescaling. Instructor-mediated discussion helps students interpret the analytical process and connect their graphical observations to empirical modeling. The instructor's role is to respond to students' evolving interpretations of the data. A study of a nontraditional laboratory section examined an instructor's efforts to shift students' framing of the activity. The study suggests that the success of such efforts depends in part on the instructor's responsiveness to students' current framing.[6] The prompts listed in Sec. S4 of the Supplementary Material are therefore examples rather than a script; which prompt is useful depends on what a group has already noticed. Other implementation details, including a typical timeline, acceptable student outcomes, and informal classroom observations, are provided in Secs. S5–S7 of the Supplementary Material.

## III. SCALING ANALYSIS AND DATA COLLAPSE

### A. Initial variable selection and qualitative reasoning

At the outset of the experiment, students conduct preliminary heating runs to become familiar with the experimental setup and to observe the thermal response. From these initial observations, students are asked to identify qualitative relationships: longer heating leads to a larger temperature increase, higher heater power produces a faster temperature increase, and a larger mass of water is more difficult to warm. Based on these correlations, students select a set of variables intended to describe the phenomenon.

A common initial choice treats the heating time $t$ as the independent variable and the temperature increase $\Delta T$ as the dependent variable, with the heater power $P$ and the mass of water $M_{\mathrm{w}}$ treated as parameters. This choice defines an initial four-dimensional variable space, $(t, \Delta T, P, M_{\mathrm{w}})$, which serves as the starting point for subsequent analysis. Students construct plots during data collection using spreadsheet software so that the structure of the data can be discussed while the experiment is still in progress. We normally encourage groups to use three heater powers and two water masses for model construction, with additional conditions used to test their model-based predictions. Further details of the implementation are provided in Sec. S3 of the Supplementary Material.

**B. First scaling step: from time to energy input**

When the raw data are plotted as $\Delta T$ versus $t$, distinct sets of points appear for different combinations of heater power and water mass (Fig. 1). Rather than acting as a direct cause of temperature increase, time tracks the progression of energy transfer from the heater to the system. The transition from time to input electrical energy $Q$ is supported through instructor-mediated questioning; for example, the instructor may ask, “If no electrical power is supplied, does the temperature increase simply because time passes?” Such questions help students distinguish time as a recorded variable from accumulated input energy as a physically relevant variable. These considerations motivate replacing heating time $t$ with the accumulated input electrical energy $Q = Pt$ as the horizontal-axis variable. Replotting the data using $Q$ as the independent variable causes the curves corresponding to different heater powers to collapse for each fixed water mass (Fig. 1), reducing the effective variable space from $(t, \Delta T, P, M_\mathrm{w})$ to $(Q, \Delta T, M_\mathrm{w})$.

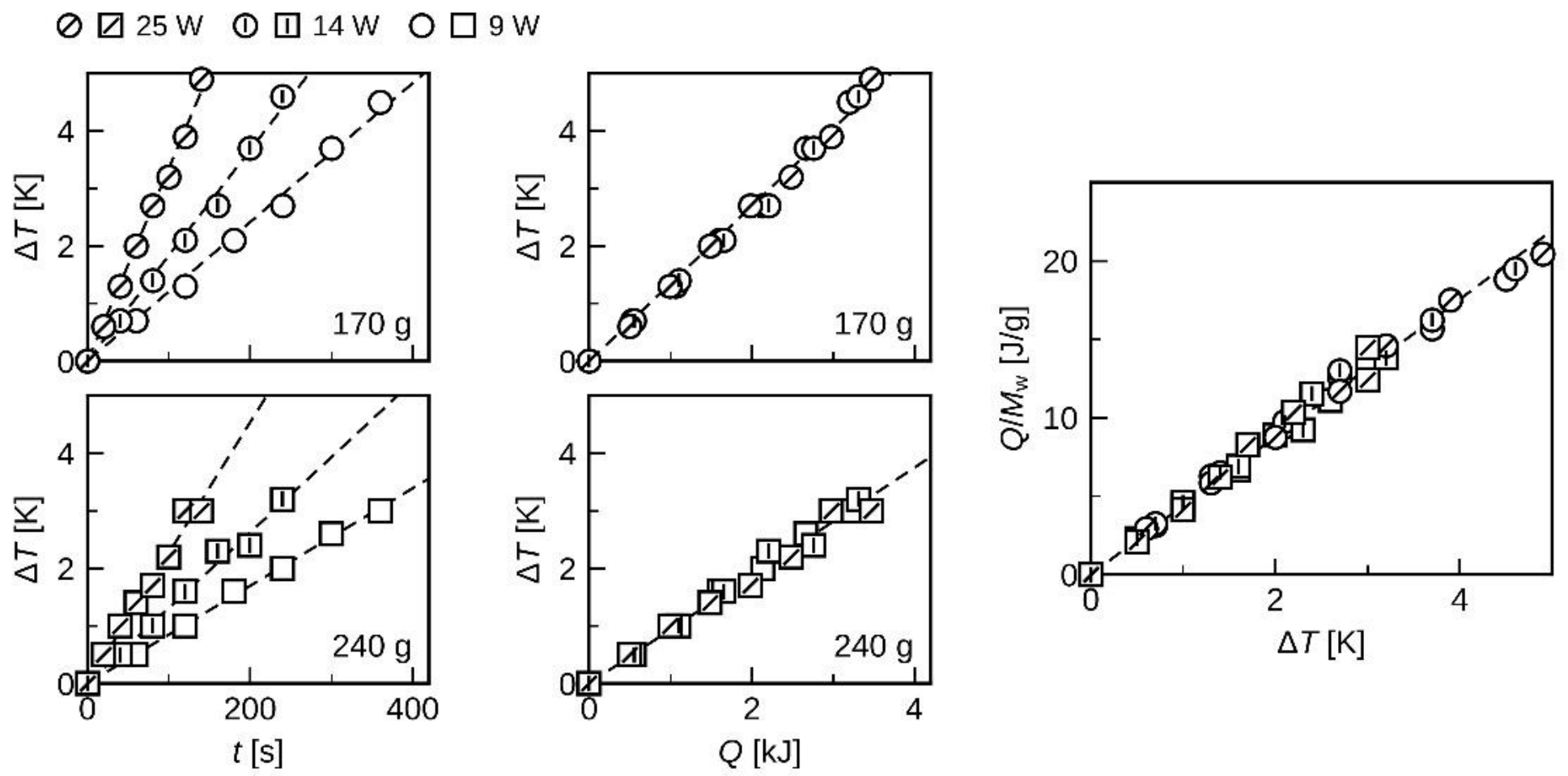

**FIG. 1.** Stepwise rescaling of the temperature increase–time data. $Q = Pt$ is the input electrical energy, $M_\mathrm{w}$ is the mass of water, and $\Delta T$ is the temperature increase. Left: $\Delta T$ plotted against heating time for different heater powers and water masses. Center: The same data plotted against $Q$. Right: $Q/M_\mathrm{w}$ plotted against $\Delta T$, showing approximate collapse onto a common straight line. Dashed lines are origin-constrained least-squares fits for each run (left), for each water mass (center), and for all data (right); the slope in the right panel is $\Gamma$ in Eq. (1). Circles and squares denote 170 g and 240 g of water; slashed, vertically barred, and open symbols denote heater powers of 25, 14, and 9 W, respectively. The left and center panels show the temperature response of the system to the input. In the right panel, the axes are interchanged so that the energy input per unit water mass required for a given temperature increase can be read directly.

### C. Second scaling step: normalization by mass and data collapse

Although the first scaling step absorbs the explicit dependence on heater power into $Q = Pt$, the data sets associated with different water masses still fall along different lines. Students are then encouraged to consider how the required energy input depends on the amount of material being heated. Rewriting the problem in terms of the energy required to achieve a given temperature increase suggests normalization by water mass.

The rescaling is guided by the proportional trend observed in the data rather than by formal dimensional analysis. The aim is to help students recognize that data obtained under different conditions can be compared by placing them on a common physical scale. When the input energy is rescaled as $Q/M_{\mathrm{w}}$, the data obtained for different water masses collapse approximately onto a common straight line in the two-variable plane $(\Delta T, Q/M_{\mathrm{w}})$ (Fig. 1). The resulting data are well described by the linear relation

$$\frac{Q}{M_{\mathrm{w}}} = \Gamma \Delta T. \qquad (1)$$

Here, $\Gamma$ is an empirical constant of proportionality with units of J $g^{-1}$ $K^{-1}$. Because $Q$ is accumulated and $\Delta T$ is measured from the start of each run ($t = 0$), both are zero at that instant by construction; the least-squares fit is therefore constrained to pass through the origin. For the model-construction data in Fig. 1, the fit gives $\Gamma = 4.38 \pm 0.04\ \mathrm{J\,g^{-1}\,K^{-1}}$, where the uncertainty is the standard error of the fit. Equation (1) constitutes an empirical model of the observed thermal response, constructed directly from the experimental data without invoking a thermodynamic formula. At this stage, $\Gamma$ simply characterizes the collapsed relation; its physical interpretation is deliberately left open. Mathematical details of the scaling procedure are provided in Sec. S1 of the Supplementary Material.

**D. Predictive test**

To assess whether the empirical model has predictive capability beyond simple curve fitting, students test it against additional data not used in its construction. When these data are plotted in the rescaled variables, they align with Eq. (1) within experimental uncertainty (Fig. 2); a fit of Eq. (1) to the 200 g test data gives a slope of $4.35 \pm 0.08\ \mathrm{J\,g^{-1}\,K^{-1}}$, consistent with $\Gamma$. This indicates that the empirical model describes data from conditions not used in its construction but still within the explored experimental domain. With Eq. (1) constructed and tested, the data-driven stage of the activity is complete. What the empirical constant $\Gamma$ represents is considered in the post-experimental discussion in Sec. IV.

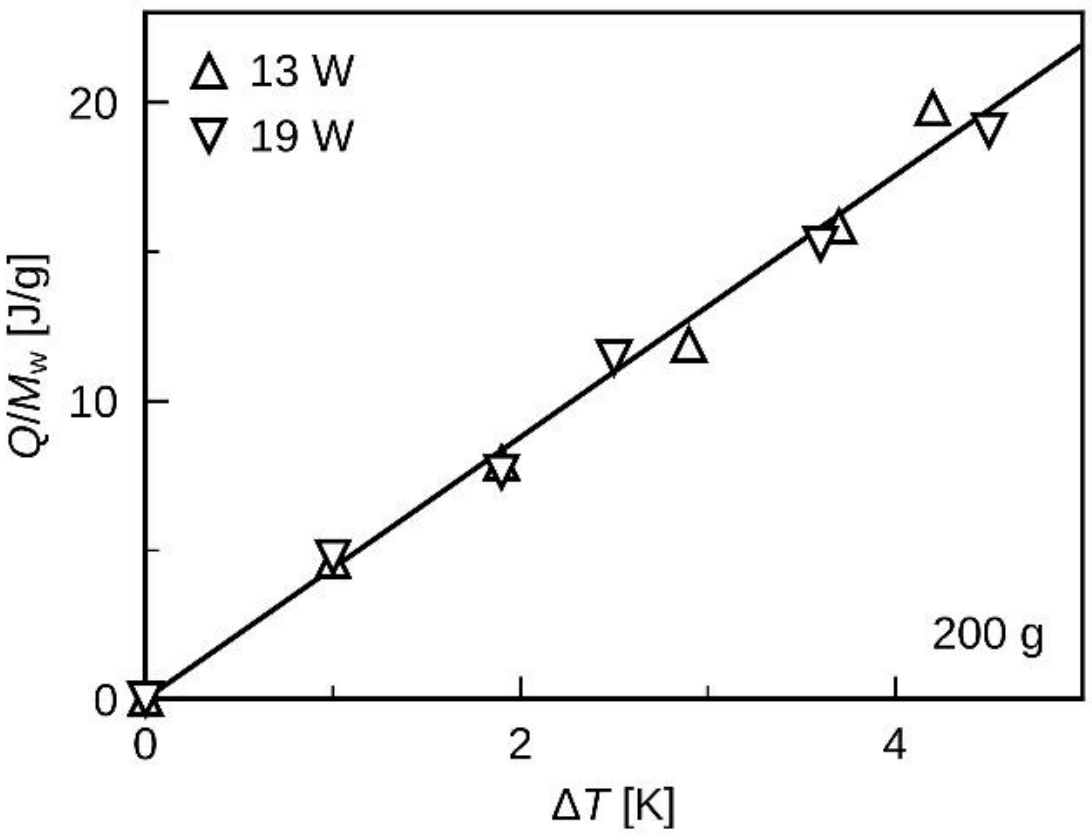

**FIG. 2.** Predictive test of the empirical model using data not included in model construction. The test data were obtained with 200 g of water at heater powers of 13 W (upward triangles) and 19 W (downward triangles). The solid line represents the prediction from Eq. (1), with $\Gamma = 4.38\ \mathrm{J\,g^{-1}\,K^{-1}}$, determined from the model-construction data. Further details of the measurement conditions are provided in Sec. S2 of the Supplementary Material.

## IV. INTRODUCTION OF THERMODYNAMICS

The post-experimental discussion begins with an observation: although the thermometer measures the temperature of the water, the calorimeter is also warmed during each run. This raises a question about the empirical parameter $\Gamma$: does it characterize the water alone or the response of the experimental system as a whole? The thermodynamic description is introduced to answer this question.

### A. Thermodynamic interpretation

The analysis assumes that the water and calorimeter undergo approximately the same temperature rise and that heat loss and energy storage in other forms are negligible. Under these assumptions, the first law of thermodynamics gives the energy balance $Q = (c_{\mathrm{w}} M_{\mathrm{w}} + C_{\mathrm{cal}})\Delta T$. Comparison with Eq. (1) then gives

$$\Gamma = c_{\mathrm{w}} + \frac{C_{\mathrm{cal}}}{M_{\mathrm{w}}}. \qquad (2)$$

Here, $c_{\mathrm{w}}$ is the specific heat of water and $C_{\mathrm{cal}}$ is the heat capacity of the calorimeter. Thus, $\Gamma$ is not a property of water alone; it characterizes the energy required per unit temperature rise of the water–calorimeter system, normalized by the water mass.

The thermodynamic description shows that the origin constraint in Eq. (1) is not an approximation. Dividing the energy balance by $M_{\mathrm{w}}$ gives, for each water mass, $Q/M_{\mathrm{w}} = (c_{\mathrm{w}} + C_{\mathrm{cal}}/M_{\mathrm{w}})\Delta T$, which passes through the origin of the $(\Delta T, Q/M_{\mathrm{w}})$ plane. In this plane, the calorimeter enters only the slope, through the term $C_{\mathrm{cal}}/M_{\mathrm{w}}$, which depends on water mass. Thus, accounting for the calorimeter does not make the origin constraint approximate; it makes the use of a single $\Gamma$ for all water masses an approximation.

Equation (2) also explains why $\Gamma$ appears approximately independent of water mass. Approximating $C_{\mathrm{cal}}$ by the heat capacity of the copper cup alone and using the tabulated specific heat of copper, we estimate $C_{\mathrm{cal}} \approx 42$ J K$^{-1}$ for the 110 g cup. The term $C_{\mathrm{cal}}/M_{\mathrm{w}}$ is therefore approximately 0.25 J g$^{-1}$ K$^{-1}$ for 170 g and 0.18 J g$^{-1}$ K$^{-1}$ for 240 g. Its predicted change over the explored mass range is only about 0.07 J g$^{-1}$ K$^{-1}$, comparable to the uncertainty, 0.08 J g$^{-1}$ K$^{-1}$, in the difference between the values obtained by fitting Eq. (1) separately to the 170 g and 240 g data. The corresponding mass dependence can therefore remain hidden within the scatter of the data in Fig. 1. Over the explored range of water masses, Eq. (1) remains an effective empirical description.

**B. Separation of the water and calorimeter contributions**

We next use the mass-dependent data to estimate $c_{\mathrm{w}}$ and $C_{\mathrm{cal}}$ without using a materials-based estimate of the calorimeter heat capacity. The mass dependence in Eq. (2) provides a means of separating the water and calorimeter contributions. Multiplying Eq. (2) by $M_{\mathrm{w}}$ and defining $K \equiv Q/\Delta T$ gives

$$K = c_{\mathrm{w}} M_{\mathrm{w}} + C_{\mathrm{cal}}. \qquad (3)$$

A plot of $K$ against water mass therefore has slope $c_{\mathrm{w}}$ and intercept $C_{\mathrm{cal}}$. In the $(M_{\mathrm{w}}, K)$ plane, the calorimeter contribution appears as the intercept, whereas in the $(\Delta T, Q/M_{\mathrm{w}})$ plane of Eq. (1), it changes only the slope through the term $C_{\mathrm{cal}}/M_{\mathrm{w}}$ (Sec. IV A).

Values of $K$ were obtained by fitting $Q$ against $\Delta T$ through the origin separately for each water mass. After serving as independent data for the predictive test in Fig. 2, the 200 g data were included in this subsequent analysis because they provide an additional water mass for determining the slope and intercept.

A weighted linear fit to the three values of $K$, shown in Fig. 3, gives $c_w = 4.56 \pm 0.24\ \mathrm{J\ g^{-1}\ K^{-1}}$ and $C_{\mathrm{cal}} = -34 \pm 45\ \mathrm{J\ K^{-1}}$. The fitted slope lies about 1.6 standard errors above the tabulated specific heat of water, 4.18 J $g^{-1}$ $K^{-1}$. Both zero and the approximately 42 J $K^{-1}$ estimate for the copper cup lie within two standard errors of the fitted intercept. The intercept is obtained by extrapolating from the smallest water mass to zero over a distance approximately 2.4 times the measured mass range. A rough estimate that also includes the immersed parts of the stirrer, heater rods, and thermometer probe is given in Sec. S2.3 of the Supplementary Material.

The quoted uncertainties are obtained from the weighted fit, treating the standard errors of the fitted values of $K$ as fixed. They do not include uncertainty associated with analysis choices or other systematic effects; the sensitivity to the selected fitting range is examined in Sec. S2.3 of the Supplementary Material, where further details of the fits are also provided.

Subtracting the fitted term $C_{\mathrm{cal}}\Delta T$ from $Q$ and normalizing by $M_{\mathrm{w}}$ places the data approximately on a common line with slope $c_{\mathrm{w}}$. The analysis therefore separates the water and calorimeter contributions while also showing the precision with which the present data determine them.

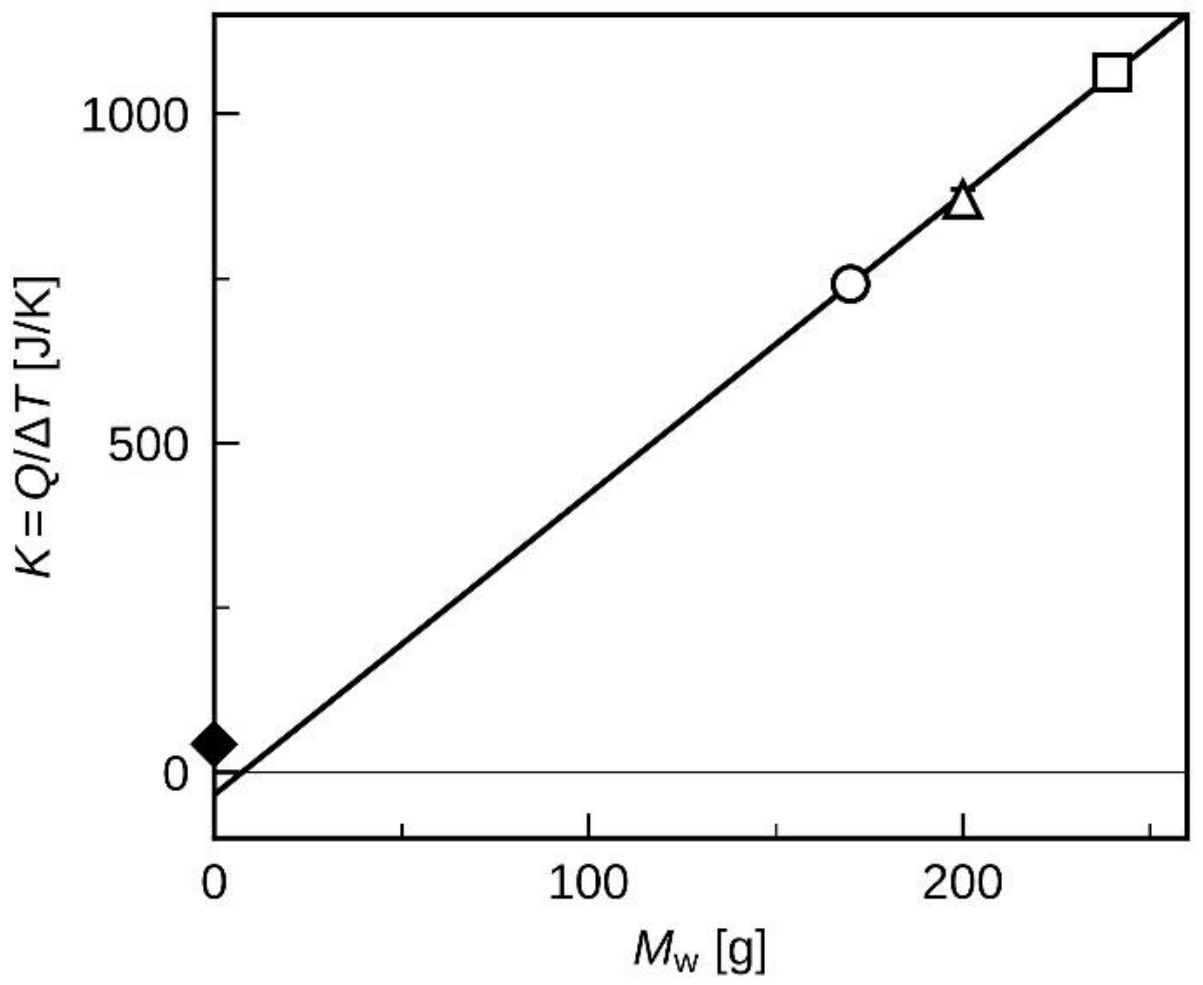

**FIG. 3.** Separation of the water and calorimeter contributions using the mass dependence of $K = Q/\Delta T$. The open circle, triangle, and square show the values of $K$ obtained from origin-constrained fits at water masses of 170, 200, and 240 g, respectively. The solid line is the weighted linear fit, whose slope gives $c_w$ and whose intercept gives $C_{cal}$. The error bars represent the standard errors of the fitted values of $K$ and are smaller than the plotted symbols. The filled diamond at $M_w = 0$ shows the estimated heat capacity of the 110 g copper cup; it is shown for comparison and was not included in the fit.

## V. PEDAGOGICAL IMPLICATIONS, DISCUSSION, AND CONCLUSIONS

We have presented a data-driven redesign of a classic specific-heat laboratory in which students construct an empirical model through successive scaling and data collapse rather than applying thermodynamic formulas from the outset. This approach requires students to decide which variables are essential and which dependencies can be absorbed through rescaling, highlighting the role of physical judgment in modeling rather than presenting modeling as a fixed procedure. In student-centered learning environments, instructor guidance is used to scaffold students' active construction of knowledge.[7] The guidance described in Sec. II B serves this role: it concerns how to proceed rather than what result to obtain.

The copper cup is central to this instructional sequence. Under the experimental conditions used here, the cup makes a physically relevant contribution, but its contribution to the mass-normalized relation changes little enough over the explored water-mass range for the data to collapse approximately onto a single relation. The thermodynamic analysis subsequently shows that the single constant $\Gamma$ approximates a quantity that depends weakly on water mass, and the mass dependence of the original data provides a basis for separating the water and calorimeter contributions. Thus, the apparatus supports three successive tasks: constructing an empirical model, interpreting the model after thermodynamics is introduced, and separating the component contributions. Students encounter these features through analysis rather than being told in advance that they are present. Treating the cup as part of the experimental system rather than merely as a source of error is one of the points the activity is designed to make. A similar difficulty arose in Fahrenheit's early mixing experiments, in which the vessel itself exchanged heat with the mixture and therefore affected the temperature being determined.[8]

The present paper describes the instructional and modeling structure of the activity but does not provide a controlled assessment of student learning outcomes. Future work could investigate students' reasoning during the transition from empirical model construction to thermodynamic interpretation and separation of the component contributions. A wider range of water masses or an independent constraint, such as one obtained from a mixing experiment, could improve the precision and stability of the separation.

## AUTHOR NOTE

## ACKNOWLEDGMENTS

This work was supported by JSPS KAKENHI Grant Number JP26K06064. The authors would like to thank Takumi Yanagihara, a former undergraduate student at Osaka Kyoiku University, for valuable assistance with the preliminary data collection and testing of the experimental setup.

## AUTHOR DECLARATIONS

### Conflict of Interest

The authors have no conflicts to disclose.

### Data Availability

The data that support the findings of this study are available from the corresponding author upon reasonable request.

**References**

1. D. Hestenes, “Toward a modeling theory of physics instruction,” Am. J. Phys. **55**, 440–454 (1987).

2. E. R. Mendoza Moreira, A. K. Palma Hernández, and G. M. Zambrano Baque, “An integrated sequence of laboratory activities to promote modelling and graphical analysis in introductory physics,” Phys. Educ. **61**, 035027 (2026).

3. E. Etkina, S. Murthy, and X. Zou, “Using introductory labs to engage students in experimental design,” Am. J. Phys. **74**, 979–986 (2006).

4. B. M. Zwickl, N. Finkelstein, and H. J. Lewandowski, “Incorporating learning goals about modeling into an upper-division physics laboratory experiment,” Am. J. Phys. **82**, 876–882 (2014).

5. C. Wieman and N. G. Holmes, “Measuring the impact of an instructional laboratory on the learning of introductory physics,” Am. J. Phys. **83**, 972–978 (2015).

6. M. Sundstrom, R. K. Fussell, A. M. Phillips, M. Akubo, S. E. Allen, D. Hammer, R. E. Scherr, and N. G. Holmes, “Instructing nontraditional physics labs: Toward responsiveness to student epistemic framing,” Phys. Rev. Phys. Educ. Res. **19**, 020140 (2023).

7. S. Hoidn, *Student-Centered Learning Environments in Higher Education Classrooms* (Palgrave Macmillan, New York, 2017), Chap. 2.

8. H. Chang, *Inventing Temperature: Measurement and Scientific Progress* (Oxford University Press, New York, 2004), pp. 225–226.

**SUPPLEMENTARY MATERIAL for**

**"Data-driven modeling in the introductory physics laboratory: Scaling analysis and data collapse in the specific heat of water experiment"**

Kazumasa Kushida[1]*, Tomohiro Oda[2], Koji Yamaguchi[3]

[1] Department of Educational Collaboration, Osaka Kyoiku University, Kashiwara, Osaka 582-8582, Japan

[2] Ikeda Senior High School Attached to Osaka Kyoiku University, Ikeda, Osaka 563-0026, Japan

[3] Tennoji Senior High School Attached to Osaka Kyoiku University, Tennoji, Osaka 543-0054, Japan

## S1. MATHEMATICAL DETAILS OF THE SCALING PROCEDURE

This section summarizes the mathematical steps underlying the scaling analysis presented in Sec. III of the main text. It describes the procedure based on the empirical patterns in the data shown in Fig. 1.

### S1.1 Initial linear fits for fixed parameters

For each combination of heater power $P_i$ and water mass $M_{\mathrm{w},j}$, the temperature increase varies approximately linearly with heating time over the restricted temperature range used in the activity. The data are therefore fitted through the origin by

$$\Delta T = \alpha_{ij} t, \qquad \text{(S1)}$$

* *Corresponding Author*: Kazumasa Kushida
* *E-mail*: kkushi@cc.osaka-kyoiku.ac.jp

where $\alpha_{ij}$ is the fitted temperature slope for the parameter set $\left(P_i, M_{\mathrm{w},j}\right)$.

### S1.2 Rescaling from time to energy input

For a fixed water mass, the fitted slopes $\alpha_{ij}$ are approximately proportional to the heater power $P_i$. This motivates replacing heating time with the input electrical energy

$$Q = P_i t = I_i V_i t. \qquad \text{(S2)}$$

Substituting Eq. (S2) into Eq. (S1) gives

$$\Delta T = \beta_j Q, \qquad \text{(S3)}$$

where the run-specific ratios $\alpha_{ij}/P_i$ are approximately independent of heater power for a fixed water mass, within the scatter of the data, and are represented by the common value $\beta_j$.

### S1.3 Normalization by water mass and collapse onto a common relation

Empirically, the product $\beta_j M_{\mathrm{w},j}$ is approximately the same for the two water masses within the scatter of the data. Normalizing the input energy by $M_{\mathrm{w},j}$ therefore brings the data onto an approximately common relation, and Eq. (S3) can be written, using the notation of the main text, as

$$\frac{Q}{M_{\mathrm{w}}} = \Gamma \Delta T, \qquad \text{(S4)}$$

with $\Gamma = 1/\beta_j M_{\mathrm{w},j}$ treated as a common empirical constant over the explored water-mass range. Equation (S4) is Eq. (1) of the main text.

Because $Q$ is accumulated and $\Delta T$ is measured from the start of each run ($t = 0$), both are zero at that instant by construction; the least-squares fit of Eq. (S4) to all model-construction data is constrained to pass through the origin. The uncertainty quoted for $\Gamma$ in Sec. III C of the main text is the standard error of this fit.

### S1.4 Scope of the scaling relations

This procedure establishes a common empirical constant $\Gamma$ over the water-mass range explored in Fig. 1. The use of a common $\Gamma$ is an approximation supported by the data in Fig. 1 rather than an assumption of exact mass independence. The scaling procedure itself does not assign a physical interpretation to $\Gamma$. Equation (2) of the main text shows that $\Gamma$ depends weakly on water mass; Sec. IV also examines this dependence using the relation between $K = Q/\Delta T$ and $M_{\mathrm{w}}$.

## S2. APPARATUS, DATA, AND ANALYSIS DETAILS

This section describes the apparatus, measurement conditions, and analysis of the data used in the main text.

### S2.1 Experimental apparatus and measurement conditions

The calorimeter (WKA, KENIS) was purchased for approximately ¥16,500 (about US$110 in 2026). In addition to the expanded-polystyrene liner around the copper cup, the underside of the lid is lined with expanded polystyrene (Fig. S1). The coiled nichrome-wire heater is supplied with the calorimeter and is mounted between two rods fixed to the lid. The rods carry the current to the heater, which is secured by screws so that its height can be adjusted. The thermometer and a ring-shaped stirrer are inserted through holes in the lid. The outer diameter of the stirrer ring is slightly smaller than the inner diameter of the copper cup, and its geometry keeps it clear of the heater. During each run, the water was stirred continuously using short vertical strokes.

The initial water temperature was approximately 18–23 °C. Water mass was measured with a digital scale (KD-187, TANITA, Tokyo, Japan) having a display resolution of 1 g. The heater was switched on, and the current was adjusted for about 2–3 s while the water was stirred. A stopwatch was started when the adjustment was completed; this moment defines $t = 0$ and the reference temperature for $\Delta T$. Current and voltage were read from the analog ammeter and voltmeter shown in Fig. S1(a); the smallest scale divisions were 0.2 A and 0.2 V, respectively. Values were recorded to the nearest 0.1 A and 0.1 V by estimating between adjacent scale marks. The recorded current and voltage for each run are listed in Table S1.

Matching the heater powers across different water masses is convenient but not required, because the recorded power for each run is incorporated into the input energy $Q = Pt$ in the analysis of Sec. III of the main text. The 200 g test runs, performed at powers not used for model construction, also fall on the same collapsed relation in Fig. 2 of the main text. In the runs reported here, the same current and voltage readings, to the reading resolution of the meters, were obtained for 170 g and 240 g at each nominal power setting (Table S1).

**S2.2 Representative data and summary of the runs used in the manuscript**

The data used in the main text comprise eight heating runs: six runs used for model construction and two additional runs used for predictive testing (Table S1). For each experimental run, the point-by-point data consist of heating time $t$, input electrical energy $Q = IVt$, and temperature increase $\Delta T$. Table S2 shows representative data from one model-construction run so that the data format and typical magnitudes can be seen.

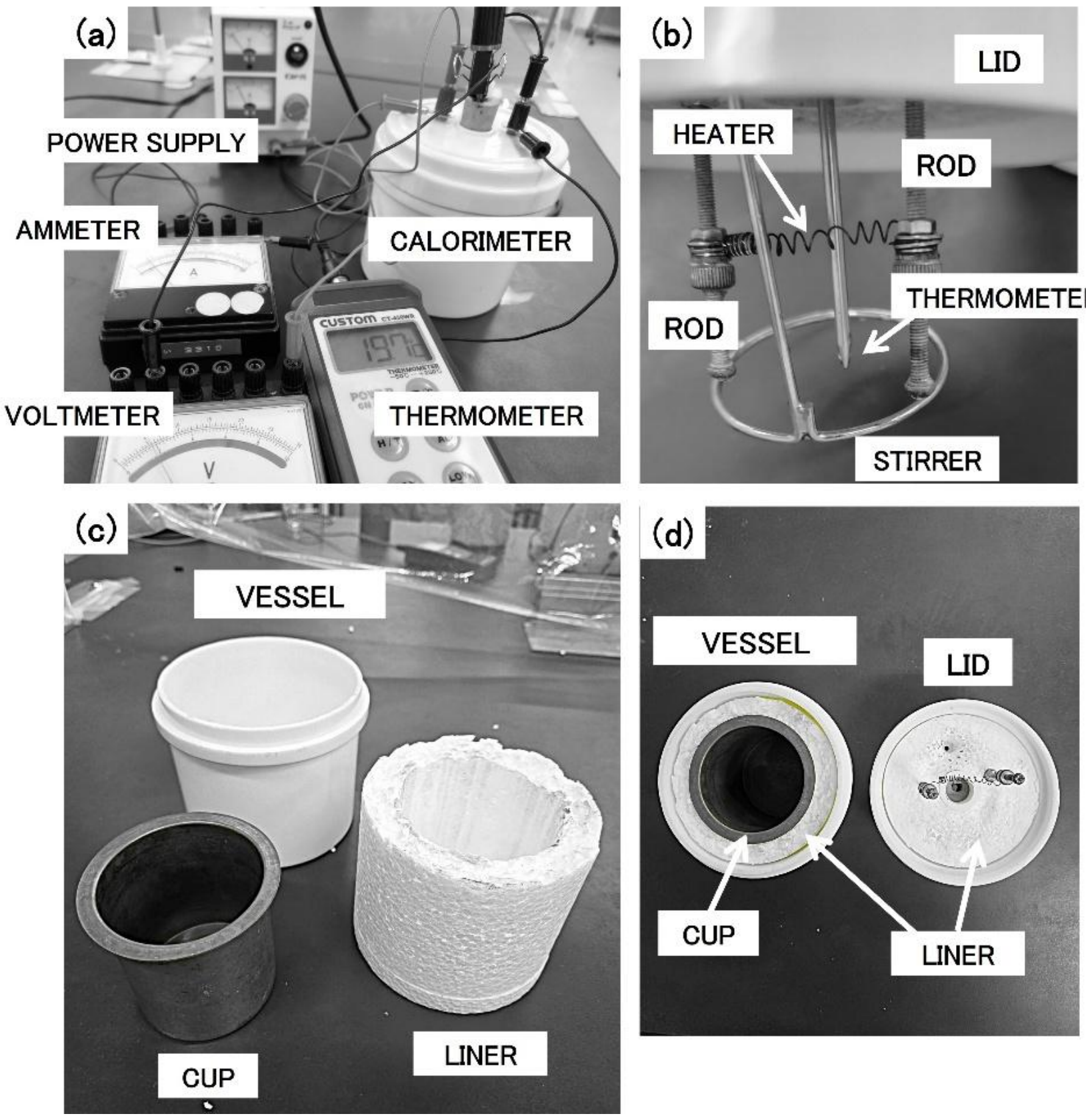

**FIG. S1.** (a) Overall experimental setup, showing the dc power supply, analog ammeter and voltmeter, digital thermometer, and calorimeter; the copper cup is enclosed by the expanded-polystyrene liner and plastic outer vessel. (b) Side view of the lid assembly, showing the coiled nichrome-wire heater mounted between two rods, the thermometer, and the ring-shaped stirrer. (c) Disassembled calorimeter components: the 110 g copper cup, the expanded-polystyrene liner, and the plastic outer vessel. (d) The assembled vessel viewed from above, together with the underside of the lid, which is lined with expanded polystyrene.

**TABLE S1.** Experimental conditions for the eight runs used for model construction and predictive testing. The same data were subsequently used in the separation analysis described in Sec. S2.3. The same current and voltage readings, to the reading resolution of the meters, were obtained for 170 g and 240 g at each nominal power setting (Sec. S2.1).

| Run ID | Role | $M_w$ [g] | $I$ [A] | $V$ [V] | $P=IV$ [W] | Power label |
|---|---|---|---|---|---|---|
| MC-170-1 | MC | 170 | 2.4 | 3.7 | 8.88 | 9 W |
| MC-170-2 | MC | 170 | 3.0 | 4.6 | 13.80 | 14 W |
| MC-170-3 | MC | 170 | 4.0 | 6.2 | 24.80 | 25 W |
| MC-240-1 | MC | 240 | 2.4 | 3.7 | 8.88 | 9 W |
| MC-240-2 | MC | 240 | 3.0 | 4.6 | 13.80 | 14 W |
| MC-240-3 | MC | 240 | 4.0 | 6.2 | 24.80 | 25 W |
| PT-200-1 | PT | 200 | 3.0 | 4.4 | 13.20 | 13 W |
| PT-200-2 | PT | 200 | 3.6 | 5.3 | 19.08 | 19 W |

MC: Model construction; PT: Predictive test.

**TABLE S2.** Representative point-by-point data from a model-construction run (MC-170-2; $M_w = 170$ g, $I = 3.0$ A, and $V = 4.6$ V). The table illustrates the data format and typical magnitudes of the quantities used in the scaling analysis.

| $t$ [s] | $Q$ [J] | $Q / M_w$ [J/g] | $\Delta T$ [K] |
|---|---|---|---|
| 0 | 0 | 0.0 | 0.0 |
| 40 | 552 | 3.2 | 0.7 |
| 80 | 1104 | 6.5 | 1.4 |
| 120 | 1656 | 9.7 | 2.1 |
| 160 | 2208 | 13.0 | 2.7 |
| 200 | 2760 | 16.2 | 3.7 |
| 240 | 3312 | 19.5 | 4.6 |

**S2.3 Fits, separation analysis, and sensitivity check**

For each water mass, $K$ was obtained from an origin-constrained least-squares fit,

$$Q = K\Delta T. \qquad \text{(S5)}$$

All runs at the same water mass were included in a common fit. The $t = 0$ points, which lie at the origin by construction, were excluded from the fits. The resulting values are $K_{170} = 741.8 \pm 7.7\ \mathrm{J\ K^{-1}}$, $K_{200} = 869.9 \pm 14.9\ \mathrm{J\ K^{-1}}$, and $K_{240} = 1062.9 \pm 15.0\ \mathrm{J\ K^{-1}}$. The quoted uncertainty in each value of $K$ is the standard error of the origin-constrained fit. After serving as independent data in the predictive test shown in Fig. 2 of the main text, the 200 g data were included in the separation analysis to provide a third water mass. Dividing $K$ by the water mass gives the mass-specific value $K/M_\mathrm{w}$. For 170 g and 240 g, the difference is

$$\frac{K_{170}}{170\ \mathrm{g}} - \frac{K_{240}}{240\ \mathrm{g}} = -0.07 \pm 0.08\ \mathrm{J\ g^{-1}\ K^{-1}},$$

as discussed in Sec. IV A of the main text.

The three values of $K$ were fitted by weighted least squares to

$$K = c_\mathrm{w} M_\mathrm{w} + C_\mathrm{cal}, \qquad \text{(S6)}$$

using $w_j = 1/\sigma_{K_j}^2$ as the weight for each mass. The fit gives $c_\mathrm{w} = 4.56 \pm 0.24\ \mathrm{J\ g^{-1}\ K^{-1}}$ and $C_\mathrm{cal} = -34 \pm 45\ \mathrm{J\ K^{-1}}$, with $\chi^2 = 0.32$ for one degree of freedom. The parameter standard errors were calculated by treating the estimated values of $\sigma_{K_j}$ as fixed. They were not rescaled by $\chi^2/\mathrm{dof}$ and do not include uncertainty in the weights, uncertainty associated with analysis choices, or other systematic effects. The fitted values of $c_\mathrm{w}$ and $C_\mathrm{cal}$ are strongly anticorrelated, with a correlation coefficient of −0.99.

The amplification of uncertainty can be seen directly in a two-mass analysis using the 170 g and 240 g model-construction data. For two masses $M_1$ and $M_2$,

$$c_{\mathrm{w}} = \frac{K_2}{M_2} + \frac{M_1}{M_2 - M_1}\left(\frac{K_2}{M_2} - \frac{K_1}{M_1}\right), \qquad \text{(S7)}$$

and

$$C_{\mathrm{cal}} = \frac{M_1 M_2}{M_2 - M_1}\left(\frac{K_1}{M_1} - \frac{K_2}{M_2}\right). \qquad \text{(S8)}$$

For $M_1 = 170$ g and $M_2 = 240$ g, the mass-dependent coefficients are $M_1/(M_2 - M_1) \approx 2.4$ and $M_1 M_2/(M_2 - M_1) \approx 580$ g. The difference $K_1/M_1 - K_2/M_2$ is comparable to its uncertainty, and because these coefficients multiply the difference, its uncertainty is amplified when $c_{\mathrm{w}}$ and $C_{\mathrm{cal}}$ are calculated. Substituting $K_{170}$ and $K_{240}$ into Eqs. (S7) and (S8) gives $c_{\mathrm{w}}^{(2)} = 4.59 \pm 0.24\ \mathrm{J\ g^{-1}\ K^{-1}}$ and $C_{\mathrm{cal}}^{(2)} = -38 \pm 45\ \mathrm{J\ K^{-1}}$, where the superscript denotes the two-mass analysis and the uncertainties were obtained by propagating the independent standard errors of $K_{170}$ and $K_{240}$. The two-mass and three-mass results differ by much less than their quoted standard errors, indicating that the inclusion of the 200 g test data does not substantially change the separation result.

Using the tabulated specific heat of copper, the heat capacity of the 110 g copper cup is estimated as $C_{\mathrm{cup}} = c_{\mathrm{Cu}} M_{\mathrm{cup}} \approx 0.385\ \mathrm{J\ g^{-1}\ K^{-1}} \times 110\ \mathrm{g} \approx 42\ \mathrm{J\ K^{-1}}$. Rough estimates based on the measured dimensions of the disassembled parts and the total mass of the stainless-steel ring stirrer indicate that the immersed portions of the stirrer, the two heater rods, and the thermometer probe add approximately 7–13 $\mathrm{J\ K^{-1}}$. The total heat capacity of the copper cup and these immersed components is therefore roughly 50–55 J K$^{-1}$. Because the immersed lengths increase with water level, part of this contribution varies with $M_\mathrm{w}$ and can enter the slope of Eq. (S6) rather than $C_\mathrm{cal}$. This range therefore indicates the scale of the apparatus contribution rather than a precise prediction of $C_\mathrm{cal}$. The fitted intercepts are consistent with zero and differ by about two standard errors from the estimated 50–55 J K$^{-1}$ scale of the apparatus contribution.

As a sensitivity check, the fit for $K_{170}$ was repeated after restricting the 170 g data to $\Delta T \leq$ 3.2 K, the range covered by the 240 g data. This restriction removes the last two points from each 170 g run, leaving 13 of the 19 points used in the original fit. The two-mass separation was then recalculated using the restricted-range value of $K_{170}$. The value of $K_{170}$ increased by about 5%, from 742 to 777 J K$^{-1}$. Because the change in $K_{170}$ is multiplied by $M_2/(M_2 - M_1) \approx 3.4$ in $C_\mathrm{cal}^{(2)}$, $C_\mathrm{cal}^{(2)}$ changed from −38 to 82 J K$^{-1}$, while $c_\mathrm{w}^{(2)}$ changed from 4.59 to 4.09 J g$^{-1}$ K$^{-1}$. These changes are approximately 2.7 and 2.1 times, respectively, the quoted standard errors of the original two-mass separation. The result is therefore sensitive to the selected fitting range, which is one reason the separation is presented as a demonstration of the thermodynamic structure rather than as a precision determination.

## S3. STUDENT INSTRUCTIONS AND ACTIVITY STRUCTURE

### S3.1 Black-box approach

Students are asked to construct an empirical model rather than begin with a textbook expression for specific heat or a thermodynamic energy-balance equation. This black-box framing does not imply unguided discovery. Students follow an activity guide that specifies the experimental procedures, parameter ranges, and data-recording methods, while instructor-mediated questioning supports the modeling process.

### S3.2 Preliminary observation, variable selection, and initial representation

Students first conduct a preliminary heating trial to become familiar with the apparatus and observe the basic features of the thermal response. With staged instructor guidance, they identify measurable quantities relevant to the phenomenon and arrive at the initial variable space $(t, \Delta T, P, M_{\mathrm{w}})$.

This stage does not provide students with a complete analytical algorithm. Rather, it supports the construction of an initial data representation that can subsequently be revised by comparing data obtained under different experimental conditions.

### S3.3 Minimal data set and experimental conditions

For constructing the empirical model, we normally encourage groups to collect data for a three-by-two set of experimental conditions, such as three heater powers and two water masses, together with two additional conditions for predictive testing.

When time is limited, a smaller data set may be used. For example, a two-by-two set of conditions can serve as a minimum workable data set for constructing the empirical relation, with one additional condition used for predictive testing when feasible.

### S3.4 Data collection and real-time plotting

Students construct plots in real time using spreadsheet software. The real-time plots allow the instructor to discuss the structure of the data with each group while the experiment is still in progress and to identify practical problems of the kind described in Sec. S4.7 before data collection is completed.

### S3.5 Scaling and data collapse

Students re-express the data first in terms of the input energy $Q = Pt$ and then in terms of the mass-normalized input energy $Q/M_{\mathrm{w}}$. Representative prompts for each step and the proportional reasoning used to motivate the rescaling are given in Secs. S4.2–S4.4.

### S3.6 Model construction and predictive test

From the collapsed data, students formulate the empirical model given by Eq. (1) of the main text and test it using additional data not included in its construction.

When time is limited, priority is given to constructing the empirical relation, and the predictive test may be omitted. The activity then proceeds to the post-experimental discussion using the group's results or, when necessary, representative data provided by the instructor.

### S3.7 Thermodynamic interpretation and separation

After the empirical model has been constructed and tested, the activity proceeds to the post-experimental discussion described in Sec. IV of the main text. With instructor guidance, students examine what the empirical constant characterizes. The thermodynamic description is then introduced, and students use the mass-dependent data to separate the water and calorimeter contributions.

## S4. INSTRUCTOR GUIDANCE AND REPRESENTATIVE PROMPTS

### S4.1 General role of the instructor

The instructor provides real-time scaffolding alongside the activity guide rather than giving a complete analytical procedure in advance. The instructor circulates among groups, examines their plots, and asks questions that help students interpret each step, reconsider their variables and representations, and construct the empirical model.

The prompts below are representative examples rather than a verbatim script. In practice, the instructor selects and adapts the prompts according to each group's progress, data quality, and prior knowledge while maintaining the overall sequence of the activity.

### S4.2 Guiding students from time to energy

The initial representation uses elapsed time as the horizontal-axis variable because time is recorded directly during the experiment. The instructor guides students to consider what physical quantity accounts for both the duration of heating and the heater power.

**Representative prompts include:**

"If no electrical power is supplied, does the temperature increase simply because time passes?"

"What quantity describes both how strongly the system is heated and how long it is heated?"

"If two runs last for the same time but use different heater powers, should they produce the same temperature increase?"

These questions help students distinguish elapsed time as a recorded variable from accumulated input energy as the physically relevant quantity and motivate the use of $Q = Pt$ as the horizontal-axis variable.

**S4.3 Guiding students from energy to mass-normalized energy**

After the data are replotted using $Q = Pt$, the data obtained at different powers for the same water mass collapse approximately onto a common relation, whereas the data for different water masses remain separated. The instructor then directs attention to the amount of water being heated.

**Representative prompts include:**

“For the same temperature increase, does a larger mass of water require more or less energy?”

“How could we compare the energy required for different amounts of water?”

“What quantity would express the energy input per unit water mass?”

These questions motivate the use of $Q/M_{\mathrm{w}}$ as a common scale for comparing data obtained with different water masses.

**S4.4 Introducing data collapse through proportional reasoning**

The scaling procedure is introduced through elementary proportional reasoning rather than formal dimensional analysis. The instructor may illustrate that proportional relations such as y = x, y = 2x, and y = 3x can be placed on a common line by rescaling one of the plotted variables. Similar reasoning can be applied when the observed slope varies approximately inversely with an experimental parameter.

**Representative prompts include:**

"If one data set changes twice as fast as another, what happens if we rescale the corresponding variable by a factor of two?"

"If the ratio of the temperature slopes resembles the ratio of the heater powers, what rescaling would remove the power dependence?"

"If the temperature slope appears to vary inversely with water mass, what rescaling would compensate for that dependence?"

The purpose is to develop the intuition that data obtained under different conditions can reveal a common empirical relation when expressed on an appropriate scale, rather than to introduce formal nondimensionalization.

**S4.5 Guiding students toward the question of what Γ characterizes**

After the empirical model has been constructed and tested, the instructor directs attention to the physical meaning of its fitted parameter.

**Representative prompts include:**

"What does the slope of the collapsed graph represent?"

"Did the heater warm only the water?"

"What other parts of the apparatus became warm?"

"Does $\Gamma$ characterize the water alone or the response of the heated system?"

"What physical description would help us determine what $\Gamma$ represents?"

**S4.6 Guiding students through the thermodynamic interpretation and separation**

**Representative prompts include:**

"What does Eq. (2) imply about the physical meaning and water-mass dependence of $\Gamma$?"

"Using the mass of the copper cup and the tabulated specific heat of copper, what is the approximate magnitude of its heat capacity?"

"How can Eq. (3) be used to determine $c_{\mathrm{w}}$ and $C_{\mathrm{cal}}$ from data obtained at different water masses?"

"Why is the intercept $C_{\mathrm{cal}}$ determined much less precisely than the slope $c_{\mathrm{w}}$?"

"Does the result change if the fits for different water masses use the same range of $\Delta T$?"

"How could the experimental design be changed to determine $C_{\mathrm{cal}}$ more precisely?"

Instructors may also note that comparison with the tabulated specific heat of water does not by itself establish the accuracy of the fitted values, because heat loss is not included in Eq. (3) and can affect both $c_{\mathrm{w}}$ and $C_{\mathrm{cal}}$.

**S4.7 Responding to experimental difficulties and incomplete progress**

In our implementation, most groups obtain data that are sufficiently linear for the purposes of the activity. Small deviations from an approximately linear trend are treated as ordinary measurement variability. Larger irregularities may result from insufficient stirring, unintended contact among the heater, thermometer, and stirrer, or poor placement of the thermometer.

The instructor first asks students to inspect their apparatus and plots and identify possible experimental causes. If necessary, the instructor suggests practical adjustments, such as repositioning the heater or thermometer or modifying the stirring motion.

With instructor scaffolding, most groups reach the stage of constructing a useful collapsed representation during the laboratory period. When time is limited, the reduced procedures described in Secs. S3.3 and S3.6 are used.

## S5. TYPICAL IMPLEMENTATION TIMELINE

The activity is implemented over two 90-minute laboratory sessions. Students work in groups of two or three. In the course described here, a class consists of approximately 30 students, and two groups normally perform the specific-heat experiment in a given week while the remaining groups work on other laboratory activities. The instructor circulates among the groups performing the specific-heat experiment approximately every 10–15 minutes.

**Elapsed time 0–90 min: Data collection**

Students become familiar with the apparatus, conduct preliminary trials, collect temperature–time data, and construct plots using spreadsheet software. In our implementation, most groups complete the data collection for model construction during this period, together with the additional runs used for predictive testing.

**Elapsed time 90–110 min: First scaling step**

Students compare their initial plots and reconsider the role of time as a plotted variable. With instructor guidance, they replace t with the accumulated input energy $Q = Pt$.

**Elapsed time 110–140 min: Second scaling step and predictive test**

Students examine the remaining dependence on water mass, introduce $Q/M_{\mathrm{w}}$, and construct the empirical relation $Q/M_{\mathrm{w}} = \Gamma\Delta T$. They then test the relation using additional data not included in its construction.

**Elapsed time 140–180 min: Post-experimental discussion**

With instructor guidance, students consider what the empirical constant characterizes. The thermodynamic description is then introduced, and students carry out the separation analysis described in Sec. IV of the main text. As time permits, they compare the fitted calorimeter heat capacity with the copper-cup estimate and consider how the precision of the separation could be improved.

## S6. ACCEPTABLE STUDENT OUTCOMES

Depending on their progress and the available time, meaningful student outcomes may include the following:

1. recognizing that elapsed time alone does not account for differences in heater power;

2. introducing $Q = Pt$ as the accumulated input energy;

3. recognizing that normalization by water mass is needed to compare data obtained with different amounts of water;

4. observing that rescaling can bring data obtained under different conditions onto an approximately common relation;

5. constructing the empirical relation $Q/M_{\mathrm{w}} = \Gamma \Delta T$;

6. testing the empirical relation using data not included in its construction;

7. recognizing that the fitted empirical constant does not by itself specify whether the measured response belongs to the water alone or to the heated system as a whole;

8. interpreting $\Gamma$ using the thermodynamic description of the water–calorimeter system;

9. recognizing that Eq. (1) remains an effective empirical description over the explored water-mass range despite the weak mass dependence introduced by the calorimeter; and

10. using the mass dependence of $K = Q/\Delta T$ to separate the water and calorimeter contributions and assess the precision and stability of the result.

For groups that do not complete all stages during the laboratory period, the remaining stages may be discussed using representative data provided by the instructor.

## S7. INFORMAL OBSERVATIONS AND SCOPE

The following observations are informal and are not presented as evidence from a controlled study of student learning outcomes. They are included only to characterize features observed during the classroom implementation.

Students sometimes express surprise when rescaling brings data obtained under different conditions onto an approximately common relation. The predictive test also provides a concrete occasion for discussing the difference between constructing a model that describes the original data and testing that model with additional data.

Students differ in whether and when they raise the question of whether the fitted empirical constant describes the water alone. Some raise it while examining the collapsed relation, whereas for others the question arises only during the post-experimental discussion. No systematic assessment of these responses was conducted.